\documentclass[letterpaper]{article}
\usepackage{spconf,amsmath}
\usepackage[pdftex]{graphicx}

\graphicspath{{./Fig/}}
\DeclareGraphicsExtensions{.pdf,.jpg,.png}

\usepackage[utf8]{inputenc}
\usepackage[T1]{fontenc}

\usepackage{color} 

\usepackage{multirow}

\usepackage{amsfonts}
\usepackage{amsmath,bm}
\usepackage{amsthm} 

\usepackage{mathrsfs}
\usepackage{scalerel}

\usepackage{url} 

\usepackage{etoolbox}

\BeforeBeginEnvironment{figure}{\vskip-2ex}
\AfterEndEnvironment{figure}{\vskip-3ex}

\newcommand{\Real}{\mathbb R}
\newcommand{\xv}{\mathrm{x}}
\newcommand{\Cm}{\mathrm{C}}
\newcommand{\Hm}{\mathrm{H}}
\newcommand{\Am}{\mathrm{A}}
\newcommand{\uv}{\mathrm{u}}
\newcommand{\pv}{\mathrm{p}}
\newcommand{\cv}{\mathrm{c}}
\newcommand{\sv}{\mathrm{s}}
\newcommand{\tv}{\mathrm{t}}
\newcommand{\nv}{\mathrm{n}}
\newcommand{\rv}{\mathrm{r}}

\title{Registration of retinal images from Public Health by minimising an error between vessels using an affine model with radial distortions}

\name{G. Noyel$^{\star \ddagger}$ \quad R. Thomas$^{\dagger}$ \quad S. Iles$^{\ast}$ \quad G. Bhakta$^{\ast}$ \quad A. Crowder$^{\ast}$ \quad D. Owens$^{\dagger}$ \quad P. Boyle$^{\star \ddagger}$}

 \address{$^{\star}$ International Prevention Research Institute, Lyon, France\\
     $^{\dagger}$ Swansea University, Swansea, Wales, United Kingdom\\
		 $^{\ast}$ DESW - Diabetic Eye Screening Wales, Cardiff, Wales, United Kingdom\\
		 $^{\ddagger}$ University of Strathclyde Institute of Global Public Health, Dardilly - Lyon Ouest, France}
\begin{document}
%
\maketitle
\begin{abstract}
In order to estimate a registration model of eye fundus images made of an affinity and two radial distortions, we introduce an estimation criterion based on an error between the vessels. In \cite{Noyel2017c}, we estimated this model by minimising the error between characteristics points. In this paper, the detected vessels are selected using the circle and ellipse equations of the overlap area boundaries deduced from our model. Our method successfully registers 96 \% of the 271 pairs in a Public Health dataset acquired mostly with different cameras. This is better than our previous method \cite{Noyel2017c} and better than three other state-of-the-art methods. On a publicly available dataset, ours still better register the images than the reference method.
\end{abstract}
%
\begin{keywords}
eye fundus images, image registration, public health, radial distortion, vessel error
\end{keywords}

%
%

\section{Introduction}
\label{sec:intro}


The existence of diabetic retinopathy (DR) screening programmes has led to the creation of large Public Health (PH) databases of colour eye fundus images which allow to perform longitudinal (i.e. temporal) analysis. This analysis is facilitated by a perfect superimposition of the images.
However, as images are often captured with different cameras with at least a year of interval, an appropriate method is necessary to correct \cite{Noyel2017c}: (i) the different positions of the patient (rotation, translation, scaling), (ii) the change of the camera (scaling and radial distortion), (iii) the radial distortion caused by the projection of the retina (a spherical cap) onto the sensor plane, (iv) the radial distortion due to the camera optics and (v) the contrast changes between the images.
For such a reason, we introduced in \cite{Noyel2017c} a two-step method which consists of a pre-processing to correct the contrast variations and a registration model  composed of an affinity and two radial distortion corrections. The model parameters are estimated with characteristic points extracted by the scale-invariant feature transform (SIFT) \cite{Lowe2004}. This estimation is generally sufficient in many images. However some may present noticeable differences on their external part, especially when the overlap area is small (less than 50 \%).
The aim  of this paper is to address this issue by using the vessels to estimate the model in addition to the SIFT points. We will provide closed-form equations of the overlap area to efficiently select the vessel parts in this area. The paper is organised in two parts. Firstly, we will present our improved method. 
Secondly, we will compare it to a recent one ``REMPE''  \cite{Hernandez2017}. We will use a PH dataset with 271 image pairs acquired mostly with different cameras \cite{Noyel2017c}. We will recall the results we obtained in \cite{Noyel2017c} in this dataset for three state-of-the-art methods. A second comparison will be performed in the publicly available dataset ``FIRE'' \cite{Hernandez_FIRE2017} associated with ``REMPE''  \cite{Hernandez2017}.
 

%

%
%

\section{Method}
\label{sec:meth}

A superimposition method requires a model of deformation and an error criterion to estimate its parameters. Let us remind the model and its estimation which were both presented in \cite{Noyel2017c}. We will then present the new error criterion and its efficient computation by selecting the vessels in the overlap area.


\subsection{The model and its estimation}

Let $\pv_1, \pv_2 \in \Real^2$ be two corresponding points in the initial images 1 and 2. Our model is based on an affine homography $\Hm$ and two radial distortion corrections $\uv_{k_1}$ and $\uv_{k_2}$:
\begin{equation}
	\uv_{k_2}(\pv_2) = \Hm [\uv_{k_1}(\pv_1)], 
	\label{eq:model}
\end{equation}
where $\Hm = \begin{bmatrix}
   \Am   & \tv \\
	 O^T & 1
\end{bmatrix}$. 
$\Am$ is a non-singular matrix of $\Real^2$ representing the linear applications (rotation, translation, scaling, etc.). The vector $\tv=[t_x,t_y]^T \in \Real^2$ is a translation.
The radial distortion correction $\uv_k$ is defined by the point $\uv_k(\pv) = [\pv^u,1]^T \in \Real^3$, in homogeneous coordinates \cite{hartley_zisserman_2004}. The undistorted point $\pv^u \in \Real^2$ is given by $\pv^u-\cv = \overline{\pv}/(1+k \left\| \overline{\pv} \right\|^2)$. 
$\overline{\pv} = \pv-\cv$ is a point whose coordinates are centred on the image centre $\cv$. If the same camera is used to capture both images, the real distortion parameters are equal to $k_1 = k_2 = k$.

\begin{figure}[!htb]
  \centerline{\includegraphics[width=0.9\linewidth]{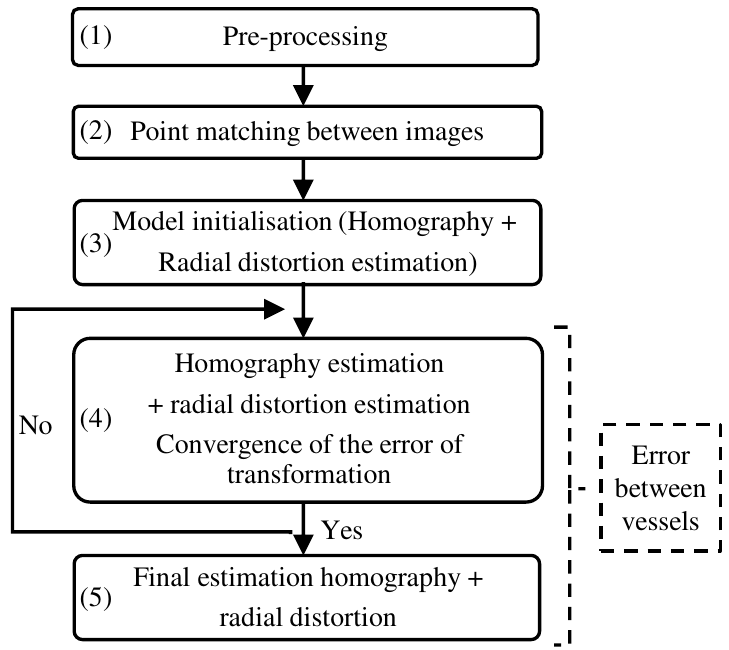}}
\caption{Flowchart of the method. A dashed brace indicates the improvement with the error between the vessels.}
\label{fig:flowchart_method}
\end{figure}

The model (Eq. \ref{eq:model}) is estimated by a several-stage approach (Fig. \ref{fig:flowchart_method}). (1) As many images in PH databases present a non-uniform brightness, a preprocessing corrects the colour contrast variations. (2) Characteristic points are extracted using SIFT algorithm \cite{Lowe2004} and matched between the images following the method presented in \cite{Noyel2017c}. (3) The matched points serve to initialise the model and the number of distortions - 1 or 2 - is automatically selected. (4) An iterative estimation is performed on the parameters until the convergence of the error. Linear estimators are used to initialise the non-linear optimisers \cite{Noyel2017c}. (5) A non-linear optimiser refines the model estimate \cite{Noyel2017c}. In this paper, at the stages (4) and (5), we will replace the SIFT-point error we used in \cite{Noyel2017c} by a criterion based on an error between the vessels.

\subsection{An error criterion between the vessels}

SIFT points are used for the initialisation stage (3) (Fig. \ref{fig:flowchart_method}). However, during the stages (4) and (5), an error between the closest vessels is minimised to ensure a better superimposition on the whole overlap area. In the plane $\Real^2$, the error is measured between the closest vessel centrelines extracted by the method of \cite{Staal2004}. Let $\mathcal{S}$ be the (centerline) curve of a source vessel and $\mathcal{R}$ the curve of the corresponding reference vessel. The curve-to-curve error $d(\mathcal{S},\mathcal{R})$ is the sum of the squared distances between all points $\sv$ of $\mathcal{S}$ to the curve $\mathcal{R}$, $d(\mathcal{S},\mathcal{R}) = \sum_{\sv \in \mathcal{S}} d^2(\sv,\mathcal{R})$ \cite{Bronstein2008}.
The squared point-to-curve distance $d^2(\sv,\mathcal{R})$ is defined for every $\sv \in \mathcal{S}$ as the squared Euclidean distance to its \textsl{closest point} $\rv^*$ on $\mathcal{R}$,
$d^2(\sv,\mathcal{R}) = \min_{\rv \in \mathcal{R}} \left\| \rv - \sv \right\|^2_{2} = \left\| \rv^* - \sv \right\|^2_{2}$. 
A second order approximation of the squared point-to-curve distance in discrete curves, was defined in \cite{Pottmann2004} by
\begin{equation}
	d^2(\sv,\mathcal{R}) \approx \frac{d}{d - \rho} [(\sv - \rv^*).\tv(\rv^*)]^2 + [(\sv - \rv^*).\nv(\rv^*)]^2
	\label{eq:SD_distance_approx}
\end{equation}
where $\tv(\rv^*)$ and $\nv(\rv^*)$ are the unit tangent and normal vectors defined at $\rv^*$. ``.'' is the scalar product. $\rho$ is the (signed) curvature radius at the point $\rv^*$. $d$ is the signed distance to the closest point $\rv^*$ defined by $d=\left\| \sv - \rv^* \right\|_2$ when $\rv^*$ and $\nv(\rv^*)$ lie on the same side of the curve and $d=-\left\| \sv - \rv^* \right\|_2$ otherwise.
In practice, the curvature radii are computed once and for all iterations before the stage (4). 
$d^2(\sv,\mathcal{R})$ is estimated between the point $s$ and its closest vessel $\mathcal{R}$. 


\subsection{Equations of the overlap area to select the vessels}

Knowing the equations of the overlap area allows to only select the vessels in this area. In each image $i \in \{1;2\}$, a circle is fitted on the boundaries of its field of view. Its centre corresponds to the image centre $\cv_i=[x_i,y_i]^T$ and its radius is denoted $r_i$.  
The circle equation of the image $i$ can be expressed by $(x-x_i)^2+(y-y_i)^2=r_i^2$ or by $\xv^T \Cm_i \xv = 0$ in matrix form, where $\xv=[x,y,1]^T$. The circle matrix $\Cm_i$ is given by:
\begin{equation}
\Cm_i = 
\begin{bmatrix}
   1 & 0 & -x_i\\
   0 & 1 & -y_i\\
	-x_i & -y_i & -r_i^2 + x_i^2 + y_i^2
\end{bmatrix}
\label{eq:matrix_circle}
\end{equation}
The radial distortion transforms the disk of radius $r_i$ into an undistorted disk of radius $r_i^u=\frac{r_i}{1+k_i r_i^2}$, whose equation is $\xv^T \Cm_i^u \xv = 0$. $\Cm_i^u$ is the same matrix as $\Cm_i$ apart from $r_i$ which is replaced by $r_i^u$. Under the homography transformation $\xv' = \Hm \xv$, the equation of the undistorted disk 1 becomes \cite{hartley_zisserman_2004}:
\begin{equation}
\xv^T \Cm_1^u \xv = \xv'^T (\Hm^{-1})^T \Cm_1^u (\Hm^{-1}) \xv' = \xv'^T \Cm_1^r \xv',
\label{eq:matrix_circle_Hrad}
\end{equation}
where $\Cm_1^r = (\Hm^{-1})^T \Cm_1^u (\Hm^{-1})$.
$\Cm_1^r$ is the matrix of a conic \cite{hartley_zisserman_2004} and in our case an ellipse. Indeed, the determinant of the $2 \times 2$ top left hand block of the matrix  is strictly positive $det(\Cm_1^r(1,2;1,2)) = (det(A))^2>0$ \cite[Chap. 7.5]{OperaMagistris}.
The overlap area between the circle $\Cm_2^u$ and the ellipse $\Cm_1^r$ is then defined by the points inside the circle and the ellipse:
\begin{equation}
\left\{
\begin{matrix}
\xv'^T \Cm_1^r \xv' &\leq& 0\\
\xv'^T \Cm_2^u \xv' &\leq& 0\\
\end{matrix}\right..
\label{eq:overlap_area}
\end{equation}
This equation system allows to only select the vessels inside the overlap area at each iteration of the model estimation.

%
%

\section{Experiments and results}
\label{sec:res}

Experiments were made in order to compare the current method to others. Two datasets were used: a Public Health dataset \cite{Noyel2017c} and a publicly available dataset. 



\subsection{Experiments in a Public Health dataset}

The PH dataset is composed of 69 randomly selected patients coming from Diabetic Eye Screening Wales (DESW) programme in the United Kingdom. All patients had diabetes and different severity stages of retinopathy or maculopathy. Each of them had been screened annually for several years and 4 images were available per screening. We selected a series of 271 image pairs: (1) of sufficient quality, (2) with an approximate screening interval of one year between the examination events \cite{Noyel2017c} and (3) captured when the screening service was renewing its eye fundus cameras. 63 \% of the pairs were captured with different cameras - different resolutions and distortions. 10 pairs had a small overlap area of about 30 \% of the superimposed image surface. All the retinal photographs were high quality and were captured according to a protocol including pupillary dilation. 
To assess the superimposition quality, all the registered overlap area were carefully checked by an expert according to the visual classification presented in \cite{Noyel2017c}. Two categories were considered: (a) no noticeable difference (i.e. \textit{correct}) and (b) noticeable difference (i.e. incorrect) with three subcategories: (b.1) differences of a small diameter vessel, (b.2) differences of the size of a large diameter vessel or (b.3) even larger. The three subcategories were grouped into a single one \textit{incorrect}.
Using this visual scale, we evaluated (i) the current method and we compared it to the results obtained in \cite{Noyel2017c} in the same dataset for three other methods: (ii) the previous one \cite{Noyel2017c} (iii) Lee et al.'s method \cite{Lee2007} and (iv) ``gdbicp'' quadratic \cite{Yang2007}. We added a comparison with a recent method (v) ``REMPE'' \cite{Hernandez2017} based on a spherical eye assumption. Standard parameters were used.


\subsection{Results in the Public Health dataset}

\begin{figure}[htb]

\begin{minipage}[b]{1.0\linewidth}
  \centering
  \centerline{\includegraphics[width=0.6\linewidth]{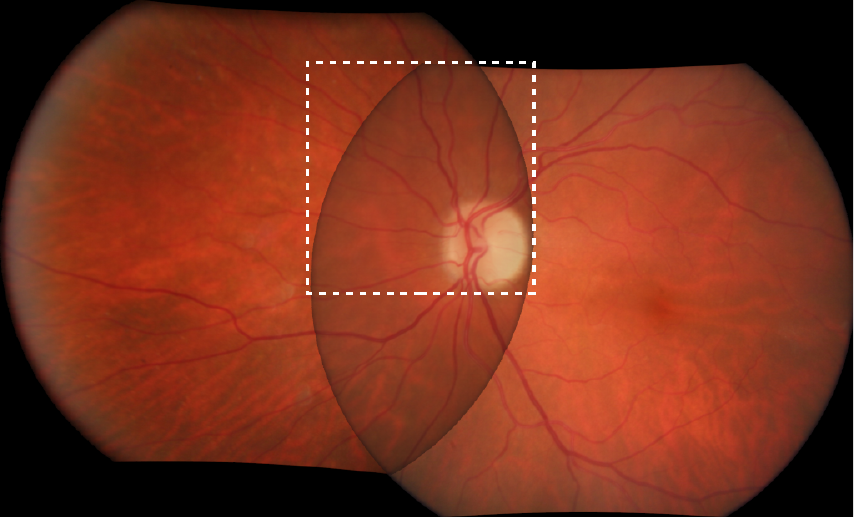}}
  \centerline{(a) Current method (i)}\medskip
\end{minipage}
\begin{minipage}[b]{0.48\linewidth}
  \centering
  \centerline{\includegraphics[width=0.6\linewidth]{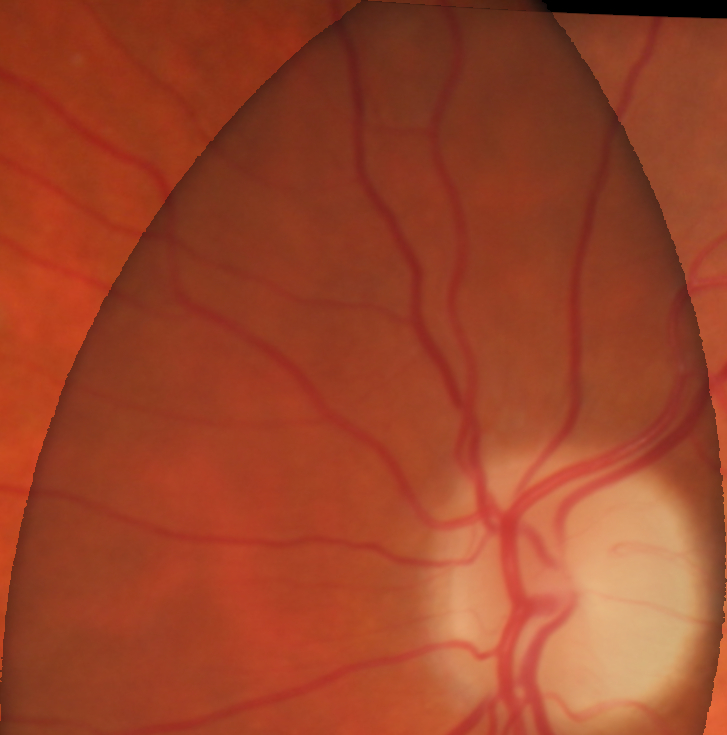}}
  \centerline{(b) Zoom of (a)}\medskip
\end{minipage}
\hfill
\begin{minipage}[b]{0.48\linewidth}
  \centering
  \centerline{\includegraphics[width=0.6\linewidth]{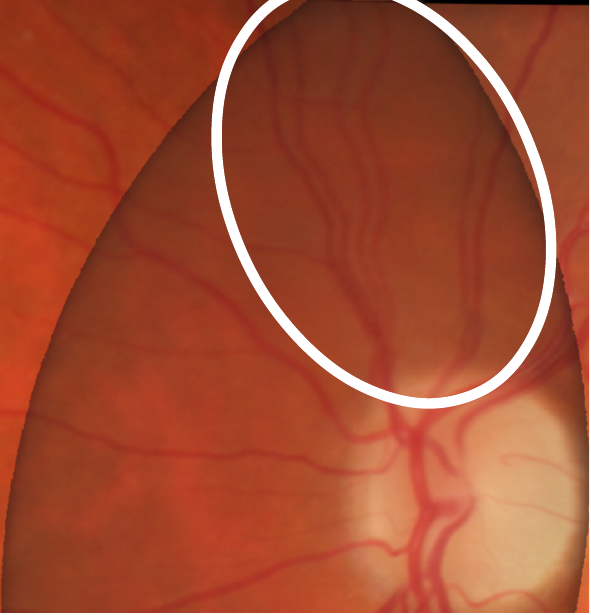}}
  \centerline{(c) Previous method (ii)}\medskip
\end{minipage}
\caption{Superimposition of a pair with a small overlap. Correct registration (a) with the current method (i). (b) Zoom of (a). (c) Incorrect registration with the  previous method (ii).}
\label{fig:res_superimpo}
\end{figure}

In table \ref{tab:res:DESW_database}, 96 \% of the pairs are correctly superimposed (i) with the current method. This methods better registers the pairs than: (ii) the previous one, (iii) Lee et al. \cite{Lee2007}, (v)  ``REMPE'' and (iv) ``gdbicp'' quadratic \cite{Yang2007}. 
Figure \ref{fig:res_superimpo} illustrates the current method (i) which successfully superimposes a pair with a small overlap, whereas the previous one (ii) fails.
Our method is therefore better than the others (ii), (iii), (iv) and (v), in this PH dataset where 63 \% of the pairs were captured with different cameras.

\begin{table}[!htb]
\begin{center}
	\begin{tabular}{@{}clc}
	\hline
	&Method 																			& \% correct \\
	\hline
	(i) & \textbf{Current model (vessels)}	      & 96 \% \\
	(ii) & Current model (SIFT-points) \cite{Noyel2017c}				& 92 \% \\
	(iii)& Lee et al. \cite{Lee2007} 							& 88 \%	\\
	(v) & ``REMPE'' \cite{Hernandez2017}					& 75 \%	 \\
	(iv) & ``gdbicp''  quadratic \cite{Yang2007}		& 74 \%\\
	\hline
	\end{tabular}
\end{center}
	\caption{Decreasing percentage of successful superimpositions for five methods in a PH dataset.}
	\label{tab:res:DESW_database}
\end{table}


\subsection{Experiments in a publicly available dataset: FIRE}

We also comparee (i) the current method to (iv) ``REMPE'' using FIRE dataset which includes a ground truth \cite{Hernandez_FIRE2017}. It is composed of 129 images forming 134 image pairs and divided into 3 categories: (1) the category $\mathcal{S}$ which contains 71 pairs with high overlap and no anatomical differences; (2) the category $\mathcal{P}$ which includes 49 pairs with small overlap and no anatomical differences and (3) the category $\mathcal{A}$ which includes 14 pairs with high overlap and large anatomical changes. The ground truth was created by experts \cite{Hernandez_FIRE2017} who selected 10 corresponding points in each image. For both methods (i) and (v), we registered these points and we computed their mean Euclidean distances between each image of the pair. 


\subsection{Results in the publicly available dataset: FIRE}

\begin{figure}[htb]

\begin{minipage}[b]{.48\linewidth}
  \centering
  \centerline{\includegraphics[width=4.0cm]{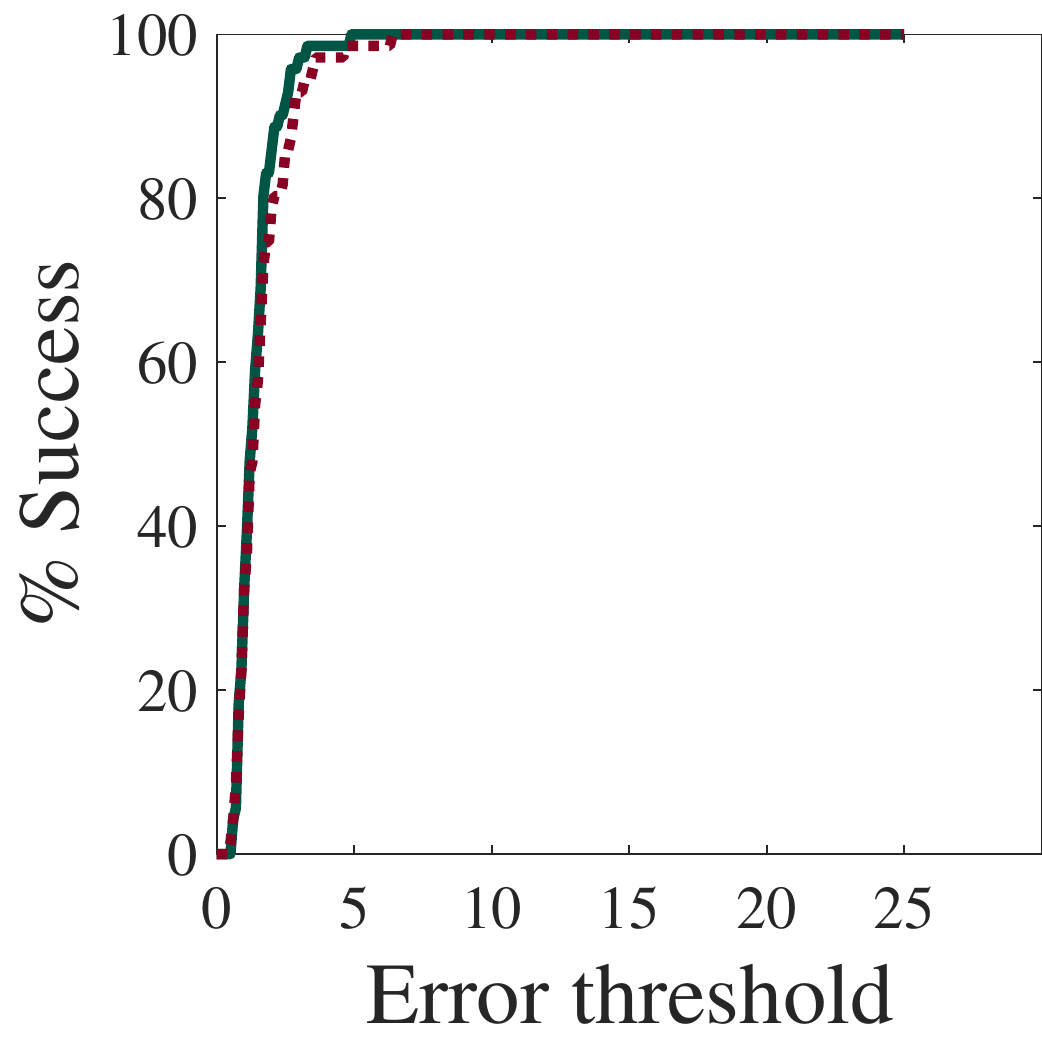}}
  \centerline{(a) Category $\mathcal{S}$}\medskip
\end{minipage}
\hfill
\begin{minipage}[b]{0.48\linewidth}
  \centering
  \centerline{\includegraphics[width=4.0cm]{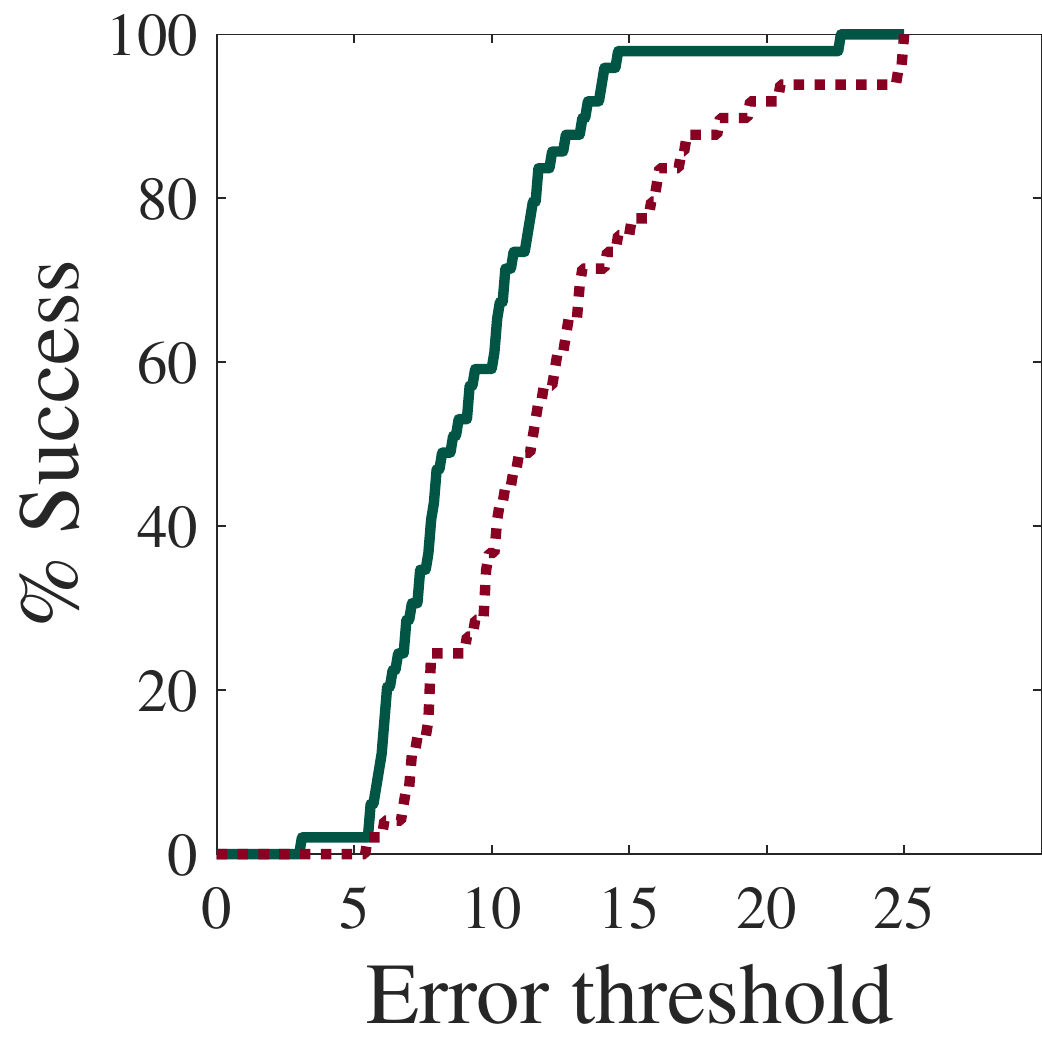}}
  \centerline{(b) Category $\mathcal{P}$}\medskip
\end{minipage}
\begin{minipage}[b]{.48\linewidth}
  \centering
  \centerline{\includegraphics[width=4.0cm]{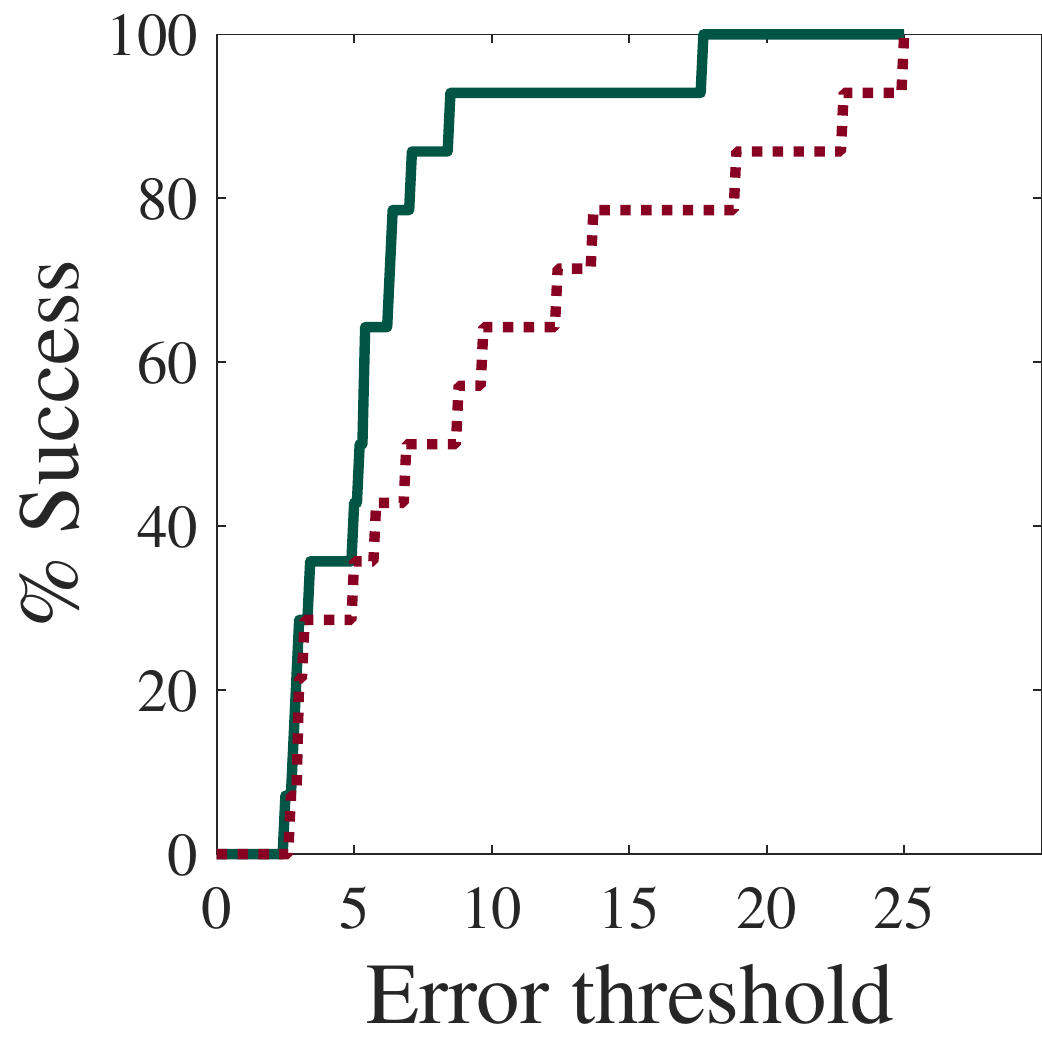}}
  \centerline{(c) Category $\mathcal{A}$}\medskip
\end{minipage}
\hfill
\begin{minipage}[b]{0.48\linewidth}
  \centering
  \centerline{\includegraphics[width=4.0cm]{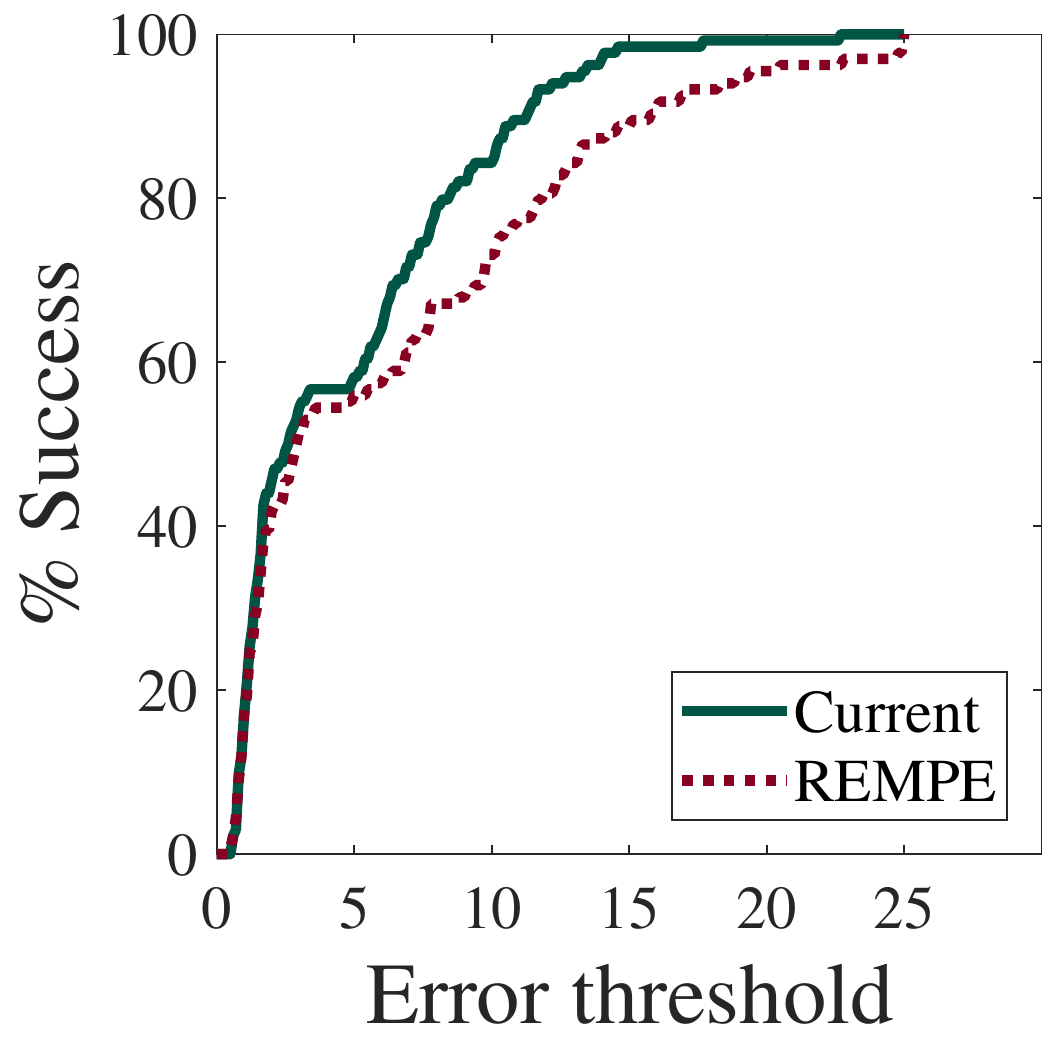}}
  \centerline{(d) Whole dataset}\medskip
\end{minipage}
\caption{Registration curves of the different categories (a), (b), (c) and the whole FIRE dataset (d). The error threshold is the value under which a registration is considered as successful. The vertical axis is the percentage of successful registrations.}
\label{fig:res:curves_FIRE}
\end{figure}

Registration error is given in the 2D plots of figure \ref{fig:res:curves_FIRE} according to the approach of \cite{Hernandez2017,Hernandez_FIRE2017}. The horizontal axis is the mean error threshold value under which a registration is considered as successful. A step of 0.1 pixels is used for the error threshold. The vertical axis is the percentage of successful registrations. The Area Under Curve (AUC) is also computed for each curve. The higher the registration curve value (or the AUC) is, the better the registration is. 
The registration curves (Fig. \ref{fig:res:curves_FIRE}), the AUC (table \ref{tab:res:FIRE_AUC}) and the mean error (table \ref{tab:res:mean_std_FIRE}) are given for each category $\mathcal{S}$ , $\mathcal{P}$ or $\mathcal{A}$ and for the whole dataset. In each case, the registration curves and the AUC of (i) the current method are higher and greater than those of (v) ``REMPE'' (table \ref{tab:res:FIRE_AUC}).  In each case, the mean error and standard deviation of (i) the current method are less than those of (v) ``REMPE''. 
The current method is therefore better than ``REMPE'' in all the categories of FIRE dataset and especially for the categories $\mathcal{P}$ (with a small overlap) and $\mathcal{A}$ (with large anatomical differences). However, for these two categories the mean error remains high, 9.25 and 5.81 pixels respectively (table \ref{tab:res:mean_std_FIRE}) for images of size $2912^2$ pixels. The superimposition can still be improved in these two categories especially for images with a small overlap (less than 50 \%).

\begin{table}[!htb]
\begin{center}
	\begin{tabular}{@{}c@{ }lcccc}
	\hline
	&Method 																	& $\mathcal{S}$ & $\mathcal{P}$ & $\mathcal{A}$ & FIRE \\
	\hline
	(i) & \textbf{Current method}	      			&	\textbf{0.942}&	\textbf{0.632}& \textbf{0.768}& \textbf{0.810}	\\
	(v) & ``REMPE''			& 0.935					& 0.511					& 0.599 				&  0.745\\
	\hline
	\end{tabular}
\end{center}
	\caption{AUC of the current method and ``REMPE'' for the categories $\mathcal{S}$, $\mathcal{P}$ and $\mathcal{A}$ and the whole FIRE dataset.}
	\label{tab:res:FIRE_AUC}
\end{table}

\begin{table}[!htb]
\begin{center}
	\begin{tabular}{@{}l@{\hspace{0.7em}}c@{\hspace{0.7em}}c@{\hspace{0.7em}}c@{\hspace{0.7em}}c@{}}
	\hline
	Method 																	& $\mathcal{S}$ & $\mathcal{P}$ & $\mathcal{A}$ & FIRE   \\
	\hline
	\textbf{Current}	      						&	\textbf{1.46} \footnotesize{(1.12)}	& \textbf{9.25} \footnotesize{(10.00)} & \textbf{5.81} \footnotesize{(7.21)} & \textbf{4.76} \footnotesize{(7.47)} \\
	``REMPE''													& 1.63 \footnotesize{(1.57)}					  & 12.64 \footnotesize{(15.19)}					& 14.05 \footnotesize{(25.73)} 			 & 6.96 \footnotesize{(13.65)}  \\
	\hline
	\end{tabular}
\end{center}
	\caption{Mean (and standard deviation) error of the current method and ``REMPE'' for the categories $\mathcal{S}$, $\mathcal{P}$ and $\mathcal{A}$ and the whole FIRE dataset. Units are in pixels.}
	\label{tab:res:mean_std_FIRE}
\end{table}

%
%

\section{Conclusion and perpsectives}
\label{sec:concl}

We have successfully achieved a new error criterion to estimate our affinity model with two radial distortions. It is based on an error between the vessels which are selected by the disk and the ellipse equations of the overlap area boundaries deduced from the model equation. Experiments have shown that our method successfully superimposes 96 \% of the pairs from a PH dataset whose images are mostly acquired with different cameras. This is better than our previous method \cite{Noyel2017c} and than three other state-of-the art methods \cite{Lee2007,Hernandez2017,Yang2007}. In the publicly available dataset, FIRE \cite{Hernandez_FIRE2017}, ours still better superimposes the images than the state-of-the-art method ``REMPE'' even if not all the pairs are perfectly superimposed. Nevertheless, the results show that our method is efficient for images of PH databases which are used for retinopathy screening and which present strong contrast variations and radial distortions.

%
%

\bibliographystyle{IEEEbib}
\bibliography{refs}

\end{document}